\documentclass[12pt]{article}

\usepackage{graphicx,amssymb,url,mathrsfs, amsmath}
\usepackage{eucal}
\usepackage{wrapfig}
\usepackage{boxedminipage}
\usepackage{setspace}
\usepackage{subfigure}

\def\l  {\lambda}

\def\IR{{\hbox{{\rm I}\kern-.2em\hbox{\rm R}}}}
\def\IB{{\hbox{{\rm I}\kern-.2em\hbox{\rm B}}}}
\def\IN{{\hbox{{\rm I}\kern-.2em\hbox{\rm N}}}}
\def\IC{\,\,{\hbox{{\rm I}\kern-.59em\hbox{\bf C}}}}
\def\IZ{{\hbox{{\rm Z}\kern-.4em\hbox{\rm Z}}}}
\def\IP{{\hbox{{\rm I}\kern-.2em\hbox{\rm P}}}}
\def\IH{{\hbox{{\rm I}\kern-.4em\hbox{\rm H}}}}
\def\ID{{\hbox{{\rm I}\kern-.2em\hbox{\rm D}}}}

\def\be{\begin{equation}}
\def\ee{\end{equation}}
\def\ba{\begin{eqnarray}}
\def\ea{\end{eqnarray}}

\begin{document}

\begin{titlepage}

\vspace{0.5in}

\begin{center}
{\large \bf {  Dense QCD: }
 a Holographic Dyonic Salt}\\
\vspace{10mm}
Mannque Rho$^{a,b}$, Sang-Jin Sin$^b$ and Ismail Zahed$^{c}$\\
\vspace{5mm}
{\it $^a$ Institut de Physique Th\'eorique, CEA Saclay, 91191 Gif-sur-Yvette, France\\
$^b$  Department of Physics, Hanyang University,  {\it 133-791} Seoul, Korea\\
      $^c$ Department of Physics and Astronomy, Stony Brook University, Stony Brook, NY 11794, USA}\\
\vspace{10mm}
  {\tt \today}
\end{center}
\begin{abstract}
Dense QCD at zero temperature with a large number of colors is a crystal.   We show that in the holographic dual
description, the crystal is made out of pairs of dyons with $e=g=\pm 1$ charges in a salt-like
arrangement.  We argue that with increasing density the dyon masses and topological charges
equalize, turning the salt-like configuration to a bcc of half-instantons.  The latter is dual to
a cubic crystal of half-skyrmions. We estimate the transition from an fcc crystal of instantons
to a bcc crystal of dyons to about 3 times nuclear matter density with a dyon binding energy
of about 180 MeV.
\end{abstract}
\end{titlepage}

\renewcommand{\thefootnote}{\arabic{footnote}}
\setcounter{footnote}{0}



\section{Introduction}

Dense QCD with a large number of colors $N_c$ is a crystal.  The Coulomb-like ratio $\Gamma=V/K$
measuring the relative interaction energy to kinetic energy is
large, ${\cal O}(N_c^2)$, since the baryon-baryon interaction is $N_c$ and the baryon kinetic energy is
$1/N_c$. Chiral skyrmions which embody key aspects of QCD at
large $N_c$~\cite{SKYRMION} crystallize~\cite{KLEBANOV} in an fcc configuration at low densities and
a cubic crystal configuration with half skyrmion symmetry at large densities~\cite{CRYSTAL,crystal2}.
The details of the crystal rearrangement and the emergence of the half-skyrmion
symmetry have been numerically explored~\cite{crystal2,park-vento}.
Some aspects of dense QCD at large $N_c$ were recently discussed in~\cite{LARRY}.

QCD at large $N_c$ and large coupling $\l =g^2N_c$ is amenable to a
holographic chiral description with baryons as flavor instantons \cite{SAKAI}.
Holographic dense QCD is expected to be a crystal of instantons, and a
preliminary analysis of the crystal structure in the Wigner-Seitz approximation
shows a transition to a free massive fermion equation of state
$n_B^{5/3}$~\cite{WIGNER}.
Here too a full scale holographic description of a dense instanton crystal
appears to be numerically involved.

In this letter we argue that some generic aspects of the crystalline structure
in holography can be elucidated using instanton physics.  In section 2 we
briefly recall the origin of flavor instantons for the description of holographic
baryons. In section 3, we show that a simple crystal arrangement of flavor
instantons is likely to split into an underlaid double lattice of BPS dyons
of opposite charges $e=g=\pm 1$ under  the action of spatial holonomies
to order $N_c\lambda$.  The arrangement is salt-like. To order
$N_c\lambda^0$ the instantons and their progeniture dyons cease to be
BPS. In section 4, we argue that with increasing density, the dyon topological
charges and masses are equalized making the final dyon rearrangement
a bcc of half instantons. The latter is dual to a cubic crystal of half-skyrmions.
In section 5 we put forward some geometrical arguments in favor of the
restoration of chiral symmetry in the holographic salt configuration.
Our conclusions and prospects are in section 6.

\section{Holographic Baryons}

Baryons in holographic dual QCD (hQCD) are sourced by instantons in bulk~\cite{SAKAI,instanton-baryon}.
The pertinent brane embedding
consists of $N_c$ D4 branes that act as a gravitational source in 10 dimensions, and
a pair of $N_f$ D8 probe branes that support color charges in the fundamental
representation~\cite{SAKAI}.  The source/probe condition is encoded in the limit
$N_c/N_f\ll 1$.  The geometry at the boundary of D4 is $R^5\times S^4$, and the ensuing
gauge theory is SYM$_{1+4}$.  By explicit compactification of $R^5\rightarrow R^4\times S^1$ with
antiperiodic boundary conditions for the boundary fermions, the boundary theory is
YM$_{1+3}$.

The open string excitations with end points on the probe branes turn
the boundary gauge theory to QCD with a mass scale defined by $S^1$ (Kaluza-Klein). This is referred to as holographic QCD (or hQCD for short).
hQCD is the dual of the boundary gauge theory in bulk as described by the sourced
gravity and the induced brane dynamics. This brane setup is characterized by a vacuum
that breaks spontaneously rigid $U(N_f)\times U(N_f)$ with massless pions (in the chiral limit) as Goldstone
bosons. Vector meson dynamics follows through the D8+$\overline{\rm D8}$ brane dynamics and is
found to follow the general lore of vector meson dominance~\cite{SAKAI,instanton-baryon,VDM}.

A single baryon in hQCD consists of a D4 wrapping around $S^4$ with a large mass $N_c\l$.
The RR flux through $S^4$ due to this wrapping is

\be
\frac 1{2\pi} \int_{S^4} \, F_4=N_c
\label{1}
\ee
which is equivalent to $N_c$ quarks. For antipodal D8+$\overline{\rm D8}$ separation along $S^1$,  the
D4 wrapping sources a flavor instanton through the coupling of the RR flux on $S^4$ with
the Chern-Simons term in $R^6$ in D8+$\overline{\rm D8}$.  The zero size instanton is BPS to leading
order. At next to leading order, the instanton size is of order $1/\sqrt{\l}$ and
follows from the minimum of the DBI action of the D8+$\overline{\rm D8}$ flavor probe
branes~\cite{instanton-baryon}.

The flavor instanton in hQCD is the core of the baryon in bulk,  that acts as a strong source in
the holographic direction, with massless pions and vector mesons strongly coupled to it. The
ensuing baryon has an electromagnetic size of order $\l^0$. The magnetic currents are vector
meson mediated. The more vector mesons, the more the currents are localized around the core,
resulting into magnetic moments which are entirely carried by the core. This  is a peculiar aspect
of the holographic baryon which is not present in the baryon of the Skyrme model~\cite{KIM}.  A
related observation regarding the holographic magnetic currents and their consequences on
the magnetic form factor of the holographic baryon was recently pointed out in~\cite{COHEN}.
Both vector meson dominance~\cite{VDM} and the Cheshire Cat principle~\cite{CHESHIRE}
emerge naturally from holography.  The dual skyrmion is just the holonomy of the flavor
instanton gauge field ${\bf A}$ along the conformal direction $Z$

\be
U(x)=P{\rm exp}\left(\int_{-\infty}^{+\infty}{\bf A}_Z(x,Z)dZ\right)
\label{2}
\ee
thereby dynamically realizing a prescient suggestion made long ago by Atiyah and Manton~\cite{MANTON}.
In leading order, the skyrmion mass is the BPS instanton mass $M=8\pi^2\kappa$ with
 $\kappa=N_c\lambda/(216\pi^3)$ in hQCD~\cite{SAKAI}.

\section{Dyonic Salt}

At low temperature, baryons are expected to crystalize at large $N_c$ irrespective
of the 't Hooft coupling $\lambda$ since the Coulomb-like ratio is
$\Gamma\approx N_c^2$. In holography, we need to consider a 3 dimensional
lattice of instantons on  $R^3\times R$ or an instanton in
 $T^3\times R$ with $R$  being the radial direction transverse
to $D4$ in  the warped Sakai-Sugimoto (SS) model~\cite{SAKAI}.
For simplicity, let us first consider an instanton in flat space.  The instanton on $T^1\times R^3$
with nontrivial holonomy was discussed in~\cite{leeyi,lee,baal,DIAKONOV}. The key observation is that the
holonomy causes the instanton to split into two electrically charged monopoles or dyons of opposite
charges. The holonomy fixes the size and charge of the monopole. Since this instanton splitting
mechanism is central to our construction below we now summarize it in flat space.

Following \cite{leeyi}, we consider a D4-D0 system with a stack of $N_f$
D4 extended along the $x^1,x^2,x^3,x^4$ with $x^4$ compactified of length $2L$.
The instanton  is D0 and is embedded in D4. The periodicity in $x^4$ makes
the $N_f=2$  instanton a caloron. We now define a T-duality along the $x^4$ direction, under which D0
becomes D1 along $x^4$  and D4 becomes D3 along $x^1,x^2,x^3$. If the holonomy along $x^4$ is
non-trivial, then D1 is stretching between two Higgsed D3's.
Since we have two ways to connect D3's, we have two monopoles
as we show in Fig.~1. Therefore the instanton  is mapped onto two monopoles
by T-duality with fractional topological charges $v$ and $1-v$ adding to 1.
These monopoles carry electric charges so they are dyons~\cite{baal}.
The masses and the contribution to the
instanton charges  of the dyons are fixed by $v$
the Higgs vev or equivalently the strength of the holonomy. For BPS instantons and therefore BPS dyons,
 the vev is arbitrary modulo  $2\pi/2L$.

\begin{figure}[]
\begin{center}
\vskip 1.5cm
\includegraphics[width=10cm,height=4.5cm]{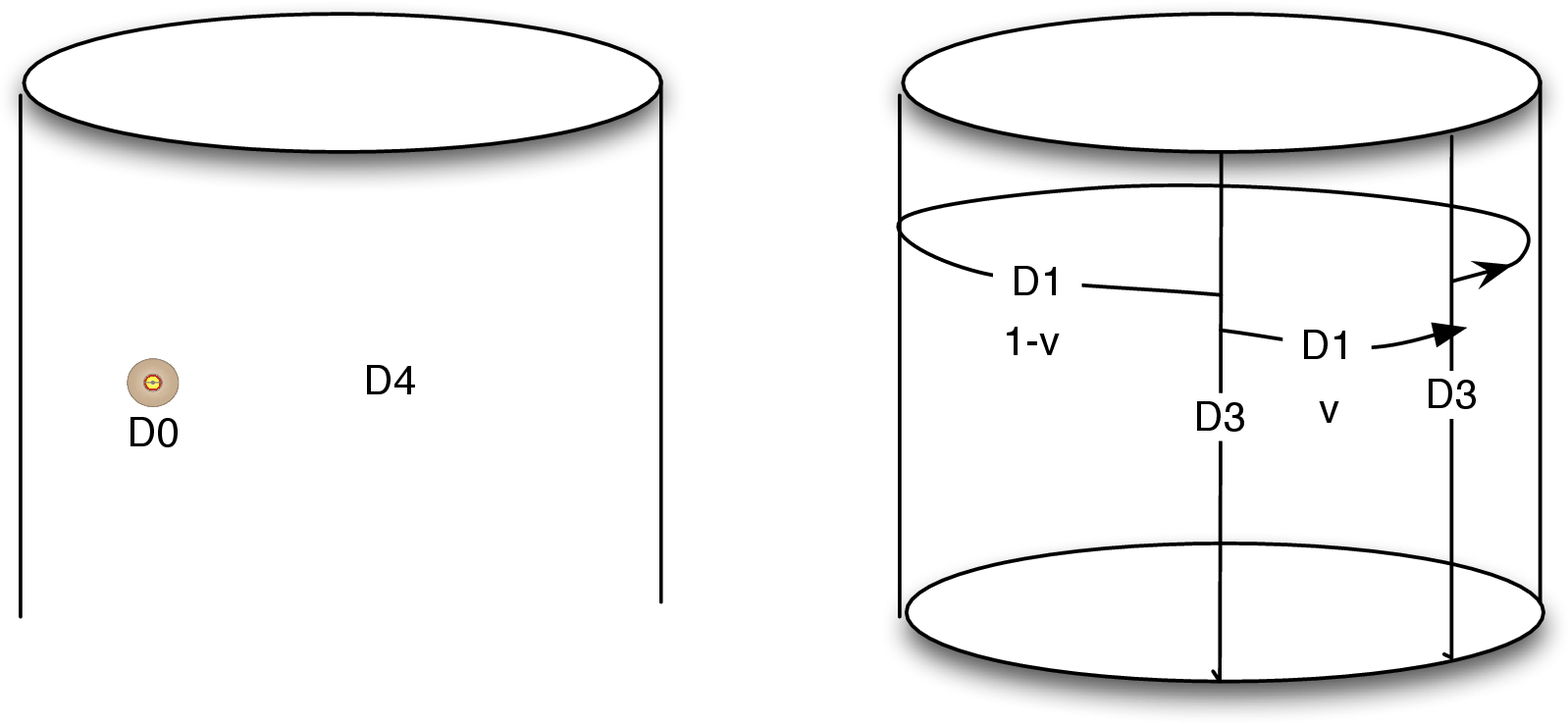}
\caption{D0 instanton in D4 without holonomy (left) and with holonomy (right) by T-duality.  The
holonomy splits the instanton into a pair of D1 monopoles. See text.}
\end{center}
\label{caloron}
\end{figure}

What will happen if we  consider a 3 dimensional array of instantons along the $x^1,x^2,x^3$ directions or an instanton on $T^3\times R$? For simplicity we set the periodicity to be $2L$ with an instanton initial arrangement in the fcc configuration.
The fcc arrangement at low density is  energetically favored over the simple cubic arrangement~\cite{park-vento}. For $N_f=2$ and by choosing the T-duality along $x^3$,
we arrive at a polarized pair of dyons of charges $e=g=\pm 1$ (in units of $T_3$)
along $x^3$. The holonomy can be chosen along $x^1$ or $x^2$ with similar
conclusions. For a fixed holonomy  in the flavor gauge field ${\bf A}_\mu$

\be
\left<{\bf A}_{3}\right>=\frac{2\pi}{2L}\,v\,T_3
\label{HOLO}
\ee-
the dyon masses are $M_+=MB_+=Mv$ and $M_-=MB_-=M(1-v)$ -- where $v$ is the Higgs vev -- with $B_\pm$ their
topological charges respectively~\cite{leeyi,lee,baal}.
We recall that $B_++B_-=v+(1-v)=1$ is the instanton
number. Here $2L$ is the cell size of the
initial fcc instanton arrangement. To order $N_c\lambda\approx \kappa$
the instantons and dyons are BPS with an arbitrary value of the vev $0\leq v<1$.  The dyonic
crystal is salt-like with intertwined lattices of topological charges $v$ and $(1-v)$ at the
vertices.  In Fig.~2 we display  the fcc instanton crystal (left) as it splits to a bcc
crystal of dyons under the action of the spatial holonomy along $x^3$.
The instantons cease to be BPS at next to leading order. We now suggest that
due to the interaction for the non-BPS objects
the vev $v$ will be fixed  to $v=1/2$ to minimize the energy, turning the dyonic salt to a bcc crystal of half-instantons.

\begin{figure}[]
\begin{center}
\vskip 0.5cm
\includegraphics[width=10cm,height=4.5cm]{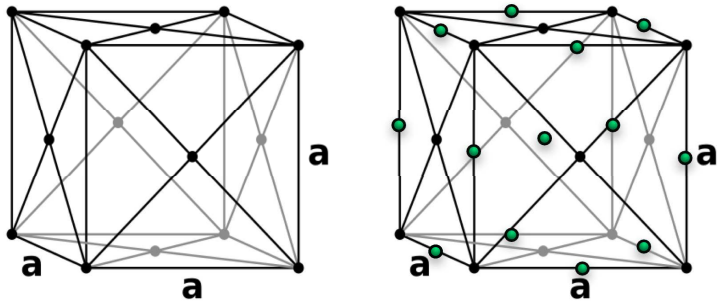}
\caption{fcc  instanton crystal at low density (left);
bcc dyon crystal at high density (right) with $a=2L$.
The spatial holonomy splits each instanton to a pair of
dyons (+,+) (black) and $(-,-$)  (green) turning the
fcc to a bcc crystal. See text.}
\end{center}
\label{LATTICE}
\end{figure}

\section{BCC Crystal of Half-Instantons}

Although the general instanton solution in warped hQCD is unknown,
to leading order in $\kappa$, the flavor instanton in hQCD is the flat space instanton with
zero size and BPS mass $M$. At next to leading order $N_c\lambda^0$ the instantons
acquire a size $\rho\approx 1/\sqrt{\lambda}$  and cease to be BPS as $M$ gets
corrected~\cite{SAKAI}.  As a result, the instantons of hQCD interact at next to next to leading
order~\cite{NN}. The core interaction is repulsive at short distances (in units of $M_{KK}$)
\be
V_\omega(r)\approx \frac{27\pi}{2}\left(\frac{N_c}{\lambda}\right)\,\frac 1{r^2}
\label{OMEGA}
\ee
and pions dominate at large distances. (\ref{OMEGA}) originates from the 4-dimensional
(topological) Coulomb repulsion. It is at the origin of the $n_B^{5/3}$ equation of state
observed in~\cite{WIGNER} for holographic instantons on $S^3$.

At  next to leading order the instantons and their progeniture dyons are non BPS
and therefore interact. The nature of the dyon interactions is in general involved~\cite{leeyi}.
Fortunately, for our dyonic crystal the details of the dyonic interactions are not important in
the non-BPS regime. Indeed, once the instantons split into $e=g=\pm 1$ dyons as in
Fig.~1, the Coulomb nature of the underlying charges will cause them to
arrange in a salt-like configuration to maximally screen the $+$ and $-$ charges, and therefore
balance the Coulomb forces.

The value $0\leq v<1$ of the holonomy is thus far arbitrary. We now note that the topological
repulsion (\ref{OMEGA}) between instantons of unit charge triggers dyons of topological
charges $v$ and $(1-v)$. The balancing of this repulsion in conjunction with the balancing of the
Coulomb forces is simply realized for $v=(1-v)$ irrespective of the details of the dyon interactions.
As a result, all dyons have equal topological charges $v=1/2$, equal masses $M/2$ and
charges $e=g=\pm 1$. The arrangement is salt-like with a unit cell $2L$ . This is a bcc crystal of
half-instantons or dyons per cubic cell $L$. The instanton or baryon density is

\be
n_B=\frac{1/2}{L^3}=\frac{4}{(2L)^3}
\label{DENS}
\ee
which is commensurate with the initial density of fcc instantons, namely a cell unit of $(2L)$
with 4 instantons. Hence our initial choice of the fcc configuration for the instantons at low
density. (\ref{DENS}) reflects on the half-instanton symmetry of the bcc dyonic salt,
which is dual to the half-skyrmion symmetry on the boundary.

To estimate the density at which the transition from the fcc crystal of instanton to the bcc crystal
of dyons occur, we note that in the flat space periodic instanton with fixed holonomy~\cite{baal,leeyi} that we shall refer to as KvLL,
the separation distance $R_{+-}$ of the dyons is~\cite{baal}

\be
R_{+-}=2\pi\,\frac{\rho^2}{2L}
\label{RPM}
\ee
with $\rho$ the KvLL instanton size with zero holonomy.
We recall that our unit cell is $2L$ for the instantons in the fcc crystal.
$v=1/2$ yields $R_{+-}=L$ which is the nearest neighbor distance in the bcc configuration.
Thus $L=\sqrt{\pi}\rho$ or a critical density $n=1/2/(\sqrt{\pi}\rho)^3$.

In hQCD, the size of the
flavor instanton is tied to its rotational inertia $\rho^2=2{\bf I}/M$~\cite{instanton-baryon,KIM}. The
moment of inertia parameter ${\bf I}$ follows by collectively quantizing the flavor instanton as a baryon.
Specifically ${\bf I}=3/2\Delta$ with $\Delta=M_\Delta-M_N$ the delta-nucleon mass splitting.
Empirically, $1/{\bf I}\approx 200$ MeV. Setting the instanton BPS mass to the nucleon mass
$M\approx 1$ GeV, it follows that $\rho\approx \sqrt{2/5}\,$ fm and $L\approx 1$
fm. So the fcc to bcc transition takes place at $n_B\approx 1/2\,{\rm fm}^{-3}$ or $n_B\approx 3\,n_{NM}$
with $n_{NM}$ the nuclear matter density. The value of $L\approx 1$ fm is amusingly close
to the numerical value of $L=1.08$ fm reported in~\cite{park-vento} for a half-skyrmion cubic crystal.

The energy density for which the fcc to bcc transition occurs can be estimated using the Madelung
constant of salt~\cite{madelung} and ignoring the contribution from the topological repulsion and the mesonic cloud
for simplicity.  The energy per instanton which is dual to the energy per baryon is $E/N=M-\Delta$,
with

\be
\Delta=(e^2+g^2)\,(T^3)^2\,\left(\int_{-\rho}^{+\rho}\frac  1{L^2+Z^2}\right)\,M_D=
(e^2+g^2)(T^3)^2\,\frac{2\,{\rm tan}^{-1}(\rho/L)}{L}\,M_D
\label{BIN}
\ee
where $\Delta$ is the energy to bring a dyon in a bcc configuration. Here
$M_D=1.748$ is Madelung constant for salt in units of the nearest neighbor
distance $L$. The Z-integration in (\ref{BIN}) is over the conformal direction
since the crystal does not extend in this direction.  It is cutoff by the KvLL instanton size $\rho$
at zero holonomy.
For $e=g=1$, $\rho=L/\sqrt{\pi}$ and $L\approx 1$ fm, the energy to assemble the core dyons in
the bcc configuration is $\Delta\approx 180$ MeV. We expect this estimate to change somehow by
including the core $\omega$ repulsion and the pion attraction, which tends to balance overall.
Our core estimate of $\Delta$ is surprisingly close to the $220$ MeV binding for the cubic crystal
of half-skyrmion symmetry in~\cite{park-vento}.

\section{Chiral Symmetry}

Does the dyonic salt configuration in bulk correspond to a chirally restored phase at high density?
This issue in the Skyrme model is rather elusive, as most numerical calculations
show only a restoration of chiral symmetry on the average per cell. In this section we suggest
a geometrical transition in bulk in support of the restoration of chiral symmetry following the splitting
of the instanton into two dyons.

In the spontaneously broken phase,  the D8+$\overline{{\rm D}8}$ configuration is depicted in Fig.~3a.
The baryon vertex corresponds to a wrapping of the D8 brane around $S^4$.  This D4 wrapping is analogous
to a point instanton in D8+$\overline{{\rm D}8}$ and is dual to a baryon on the boundary.
The baryon vertex is attached by $N_c$ strings.  Pions as Goldstone
modes are fluctuations of the holonomy along the curve $C$ as shown also in Fig.~3a.  The bare skyrmion with no meson cloud
is just the holonomy running through the core instanton.  The cloud corresponds to fluctuations on the flavor brane and tied to the core~\cite{KIM}.

Geometry alone makes it plausible that when the instanton splits into
two dyons, the geometry of Fig.~3a is replaced by the geometry of Fig.~3b, whereby the D8 and $\overline{\rm D 8}$ branes separates from each other as we explicit in Fig.~4  for a single instanton.
As a result, the baryon vertex fractionates with $N_c/2$ strings joining D8 and $N_c/2$ strings joining $\overline{{\rm D}8}$
as in Fig.~4b.  This equal opportunity splitting of
D4 corresponds to two half instantons.

The essential question here is what happens to compact D4 in D8, or the instanton wrapping $S^4$?
It is essentially  similar to what happens to D0 in D4 for the caloron in Fg.~\ref{caloron} as we discussed earlier.
To understand this, we recall that near the D8's, $S^4$ is compactified and can be ignored. Thus D8 is essentially
D4 stretching in  $x^1,x^2,x^3$ and the conformal direction $Z$ direction, while the baryon vertex is D0.
Notice that D8 is along $x^4$  near the tip therefore the $Z$ direction is a point along the $x^4$ direction.
Taking the T-dual transform along $x^4$ turns D0 into D1 and D4 into D3.
Summarizing: we obtain the D3-D1 configuration by taking the T-duality along the $x^4$ direction,
thereby explaining the splitting mechanism into dyons as shown in Fig.~4c. Throughout, the compact $S^4$
is a spectator.

Under the splitting of D8 and $\overline{{\rm D}8}$, the right and left Wilson lines decouple
\be
U^R_{1/2}(x)= P\exp\left(i\int_0^{+\infty} \,{\bf A}_Z(x,Z)\,dZ\right), \quad\quad
U^L_{1/2}(x)=P\exp \left(i\int^0_{-\infty}\, {\bf A}_Z(x,Z)\,dZ\right) \\
\ee
with each sourced by $N_c/2$ string. In each of the probe brane, the half instantons or dyons  are
repulsive through their leading core $\omega$ repulsion much like their mother instantons. They  form two distinct
$L,R$ crystals. On the boundary which is our world, the two half-instanton crystals superimpose in a maximally
repulsive configuration which is our bcc dyonic salt of half-instantons. We note that the $L,R$ crystalline structures are commensurate
with the $e=g=\pm 1$ dyonic structures as both are interchangeable by parity.

The current geometrical description is plausible but clearly not rigorous and awaits more detailed computational analysis in dense hQCD. The transition from the confined geometry to the deconfined geometry with half instantons is perhaps of the type discussed in~\cite{CHICAGO} whereby the coupling constant of the compactified boundary gauge theory is dialed to decrease in analogy with the running of the QCD coupling at high density.

\begin{figure}[]
\begin{center}
\includegraphics[width=9cm,angle=-90]{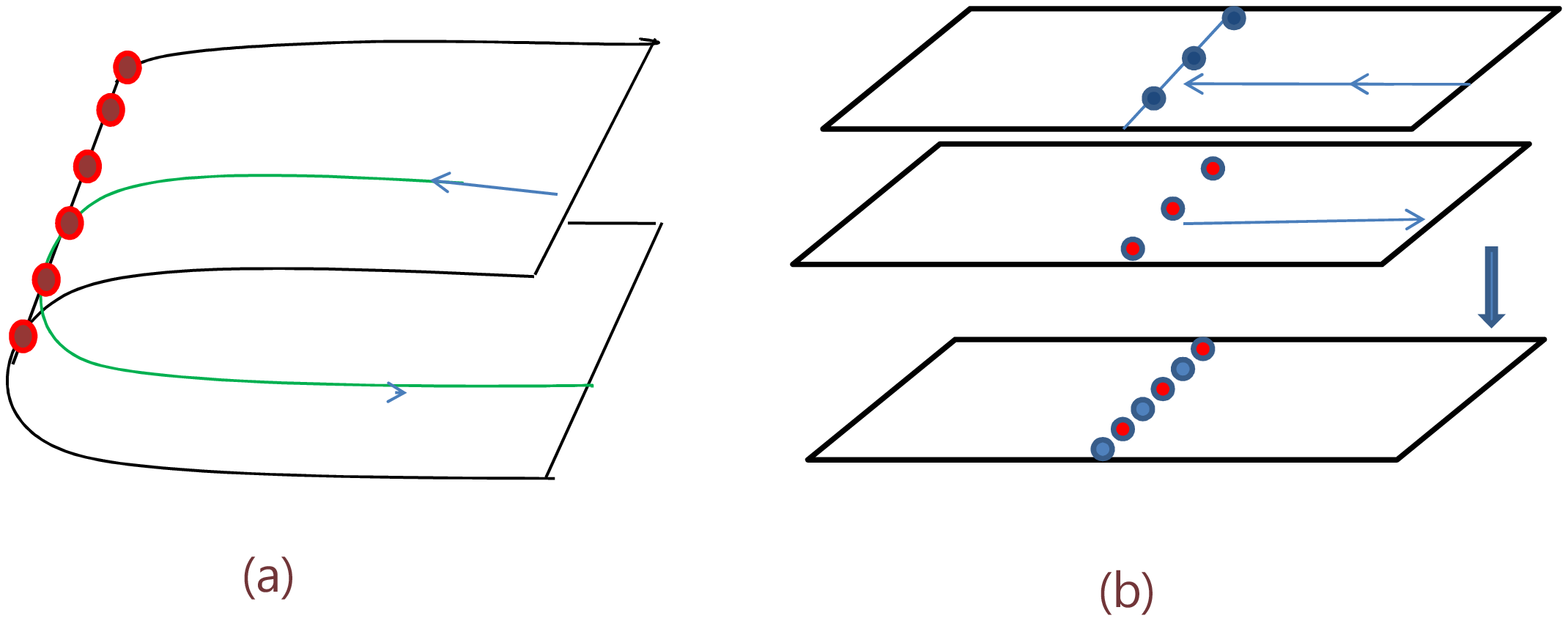}
\vskip -3cm
\caption{Dense baryonic matter: (a) low density and (b) high density. See text.}
\end{center}
\label{DB1}
\end{figure}

\begin{figure}[]
\begin{center}
\includegraphics[width=9cm,angle=-90]{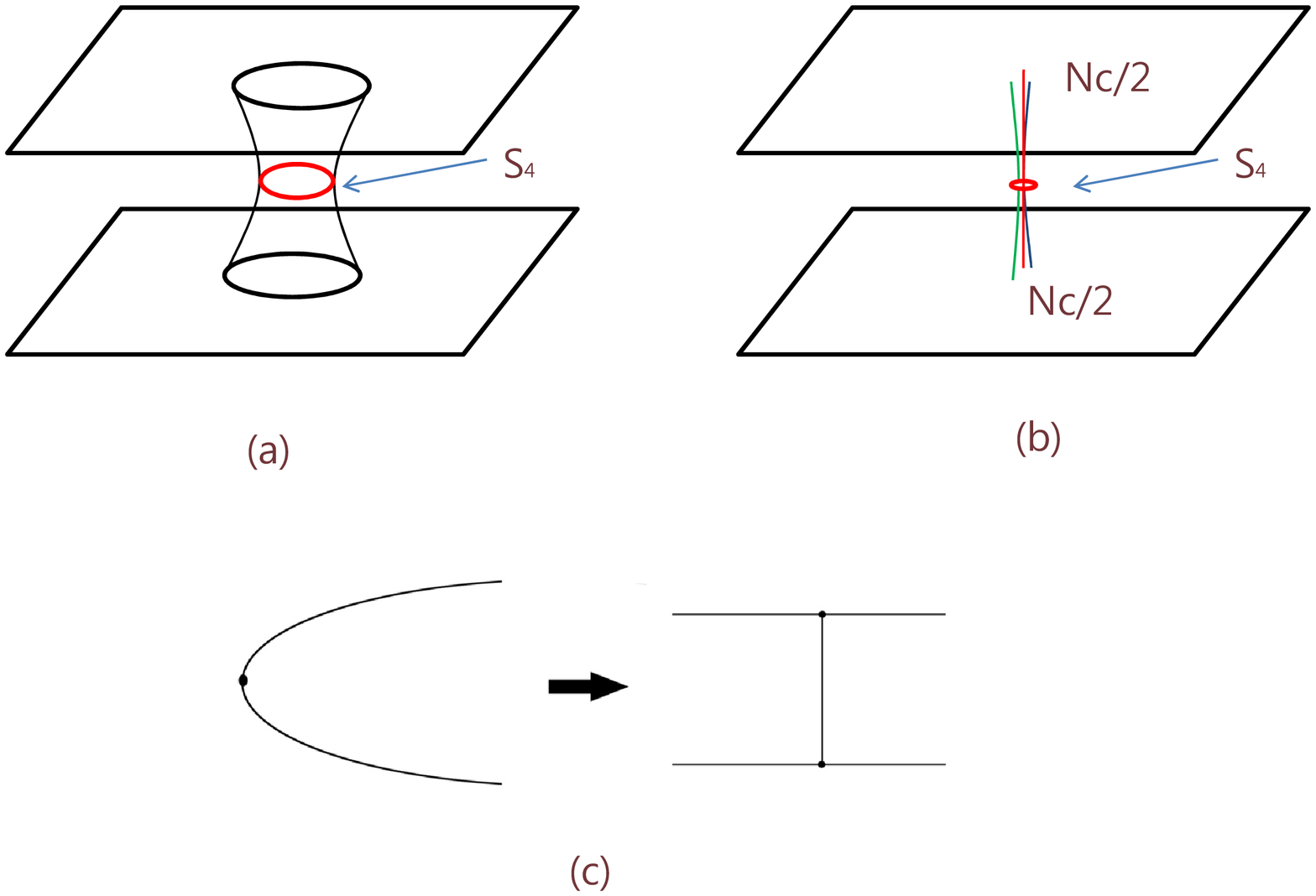}
\end{center}
\vskip -1cm
\caption{Fractionalization of the bulk instanton into two dyons. See text.}
\label{DB2}
\end{figure}

 \section{Conclusions}

In \cite{LR2}, an analogy was drawn between the half-skyrmion phase in dense baryonic matter and the half-skyrmion phase in the N\'eel magnet-VBS paramagnet transition in (2+1) D condensed matter physics~\cite{senthil}. This analogy could be better illustrated in terms of the instanton-dyon transition proposed in this paper. In the condensed matter case, the half-skyrmions confined in a skyrmion split into their constituents by the suppression of the ``monopoles" (i.e., the hedgehogs in 3D) of the emergent $CP^1$ gauge field, giving rise to a non-Ginzburg-Landau-Wilson phase that intervenes between the initial and final phases.  In holography, the half-instantons or dyons confined into the original instanton split at higher density due to the emergence of strong holonomies in the spatial directions.

We have presented general arguments for why an fcc initial arrangement of instantons should split into a bcc crystal of half instantons
or dyons. The Coulomb electric and magnetic forces between the dyons cause them to reorganize in a salt like configuration naturally
for maximum screening. The net repulsive topological interaction between the individual dyons balance naturally when all dyons in
the salt like configuration carry equal topological charges, thus equal masses.  The fcc initial arrangement, while optimal for the dyon
splitting, is not necessary in general. Indeed, once a crystal is formed at very low density, say an fcc for instance, then by increasing
the density we increase the strength of the spatial holonomies thereby causing the instanton to fractionate. The salt like configuration of equal dyons follows by balancing the Coulomb and topological forces. Conversely, the bcc configuration at high
density smoothly converts to an fcc configuration at lower density through the matching of the densities as in (\ref{DENS}) making it
energetically the most favorable.

Using the KvLL instanton to assess the dyon
separation in an instanton with a finite holonomy, we have shown that the transition from an fcc to a bcc configuration of dyons is
expected when the dyons are about 1 fm apart, or for baryon densities of the order of 3 times nuclear matter density.
This transition is also supported by a geometrical transition in bulk whereby the D4 baryon vertex on $S^4$ fractionates by $Z_2$.  We have
presented plausible geometrical arguments in favor of the restoration of chiral symmetry upon the formation of the half-instanton
bcc crystal. The latter is dual to the simple cubic crystal of half-skyrmions.

While the arguments in favor of a skyrmion crystal of half-skyrmion symmetry are not new, we believe that our holographic
accounting of this phenomenon in bulk is. Moreover, the geometrical character of the transition we have suggested together
with the physical nature of the splitting of instantons into dyons  provides the robustness needed for a topological phase
transition in QCD at large $N_c$.  Our simple estimates for the density and energy per baryon for the occurrence of the bcc
dyonic crystal are surprisingly close to those obtained using detailed numerical calculations with skyrmions~\cite{park-vento}.

Most of our estimates involve the physics of the baryon cores which is uniquely described by instantons in holography. This
is in contrast to the Skyrme model, where the core physics is at the mercy of the choice of the stabilizing term (say the fourth
order derivative term for instance) which is not fixed by current algebra. As we mentioned earlier, we expect the core estimates to be
altered somehow by the addition of the cloud contributions~\cite{KIM} as they are likely to overall balance. The current arguments
complement the instanton holographic transition reported on $S^3$~\cite{WIGNER}.

We should stress that the phenomenon described in this paper does not address color deconfinement of QCD. The half-instantons we have described are hadrons in color-confined phase, with, however, chiral symmetry restored as is the case with the half-skyrmions in cc~\cite{crystal2,park-vento}. As suggested in \cite{LR2},  this could be identified with the quarkyonic phase conjectured for large $N_c$ QCD~\cite{LARRY}.

Finally, the geometrical arguments we have presented as well as our dyonic salt configuration may be of relevance
to the finite temperature confinement deconfinement transition in QCD with colored instead of flavored
instantons and antiinstantons~\cite{DIAKONOV,ARIEL}.
While at low temperature QCD may be viewed as an instanton-antiinstanton (caloron-anticaloron) liquid, it was suggested
in~\cite{DIAKONOV} that for a critical temperature the (KvLL) instantons and antinstantons
may ionize to BPS dyons. The ionic phase may be a plasma of dyons for sufficiently
weak gauge coupling or high temperature~\cite{DIAKONOV,ARIEL,MAXIM}.
However, near the critical temperature, the gauge coupling is still strong enough
to warrant a liquid or perhaps even a weakly crystallized forms of dyons. We will return to some of these issues next.

\section{Acknowledgments}
IZ thanks Hanyang University for their kind hospitality.
This work was supported by the WCU project of Korean Ministry of Education, Science and Technology (R33-2008-000-10087-0).
IZ was supported also in part by US-DOE grants DE-FG02-88ER40388 and DE-FG03-97ER4014.

\end{document}